\documentclass{aa}
\usepackage{graphicx}
\usepackage{natbib}
\usepackage{txfonts}
\bibpunct{(}{)}{;}{a}{}{,}    % to follow the A&A style

\def\apjs{ApJS}
        
\def\jgr{J. Geophys. Res.}
          
\def\grl{Geophys. Res. Lett.}

\def\aap{Astron. Astrophys.}

\def\apj{Astrophys. J.}

\def\solphys{Sol. Phys.}
     
\def\nat{Nature}

\def\mnras{Mon. Not. R. Astron. Soc.}

\begin{document}
\title{A new approach to long-term reconstruction of the solar irradiance leads to large historical solar forcing} 
\author{A. I. Shapiro \inst{1}  \and W. Schmutz \inst{1} \and E. Rozanov \inst{1,2} \and M. Schoell \inst{1,3} \and M. Haberreiter \inst{1} \and A. V. Shapiro \inst{1,2} \and S. Nyeki\inst{1}}
\offprints{A.I. Shapiro}

\institute{Physikalisch-Meteorologishes Observatorium Davos, World Radiation Center, 7260 Davos Dorf, Switzerland\\
\email{alexander.shapiro@pmodwrc.ch}
\and Institute for Atmospheric and Climate science ETH, Zurich, Switzerland
\and Institute for Astronomy ETH, Zurich, Switzerland\\}
\date{Received 19 November 2010; accepted 22 February 2011}

\abstract
% context heading  (optional}
{The variable Sun is the most likely candidate for natural forcing of past climate change on time scales of 50 to 1000 years. Evidence for this understanding is that the terrestrial climate correlates positively with solar activity. During the past 10Õ000 years, the Sun has experienced substantial variations in activity and there have been numerous attempts to reconstruct solar irradiance. While there is general agreement on how solar forcing varied during the last several hundred years --- all reconstructions are proportional to the solar activity --- there is scientific controversy on the magnitude of solar forcing.}
% aims heading (mandatory)
{  { We present a reconstruction of the Total and Spectral Solar Irradiance covering 130 nm--10 $\mathrm{\mu}$m from 1610 to the present with annual resolution and for the Holocene with 22-year resolution.} }       
% methods heading (mandatory)
{ {We assume that the minimum state of the quiet Sun in time corresponds to the observed quietest area on the present Sun. Then we use available long-term proxies of the solar activity, which are $^{10}$Be isotope concentrations in ice cores and 22-year smoothed neutron monitor data, to interpolate between the present  quiet Sun and the minimum state of the quiet Sun. This determines the long-term trend in the solar variability which is then superposed with the 11-year activity cycle calculated from the sunspot number.}
The time-dependent solar spectral irradiance from about 7000 BC to the present is then derived using a state-of-the-art radiation code.}
% results heading (mandatory)
{ {We derive} a total and spectral solar irradiance that was substantially lower during the Maunder minimum than observed today. The difference is remarkably larger than other estimations published in the recent literature. The magnitude of the  solar UV variability, which indirectly affects climate is also found to exceed previous estimates. { We discuss in details  the assumptions which leaded us to this conclusion.}}
{}


%\titlerunning{New reconstruction technique reveals large historical variability in solar radiative forcing}
\titlerunning{A new approach to long-term reconstruction of the solar irradiance leads to large historical solar forcing}

\maketitle
%
%________________________________________________________________
\section{Introduction}\label{sec:intro}
The Sun is a variable star whose activity varies over time-scales ranging from minutes to millennia. {  Over the last thirty years the solar irradiance was measured by numerous space missions. Measurements of the Total Solar Irradiance (TSI) became available with the launch of the NIMBUS 7 mission in 1978 \citep{hoytetal1992}. Since then TSI was measured by several consecutive instruments. Each of them suffered from degradation and individual systematic effects, so the direct comparison of the measurements is impossible. Three TSI  composite based on the available data were constructed by three groups:  PMOD \citep{frohlich2006}, ACRIM \citep{willsonandmordvinov2003}, and IRMB \citep{dewitteetal2004}. These composites give quite different values of TSI, especially before 1980 and during the so-called ACRIM gap between June 1989 (end of ACRIM I observations) and October 1991 (beginning of ACRIM II observations). The most striking detail is the increase of  TSI between the minima 1986 and 1996   in the ACRIM composite and  the absence of such increase in the PMOD and IRMB composites.
Although significant progress was made during the last few years and the increase of the TSI in the ACRIM composite was strongly criticized \citep{frohlich2009,krivovaetal2009c} ,   the question of construction of the unique self-consistent TSI composite for the satellite epoch is still open. 

Measurements of the Spectral Solar Irradiance (SSI) are even more difficult and instrumental problems prevent the construction of essential composites \citep{krivovaetal2009, domingoetal2009}. Recently \citep{krivovaetal2009d} the theoretical SATIRE (Spectral And Total Irradiance REconstruction) \citep{krivovaetal2003, krivovasolanki2008} model and  SUSIM (Solar Ultraviolet Spectral Irradiance Monitor) measurements were used to reconstruct the solar UV irradiance back to 1974.

Taking into account  the problems of the reconstruction of TSI and SSI during the recent period of satellite observations and combining it with the fact that the solar dynamo, which is believed to drive all activity manifestations, is still not fully understood {\citep{dynamoLR}}, one recognizes the difficulty of reconstructing TSI and SSI to the past, when no direct measurements were available.}

Long-term changes in solar irradiance were suspected as early as  the mid-nineteenth century \citep{smyth1856}. One of the first quantitative estimates of its magnitude as well as past solar irradiance reconstructions was obtained by using the observations of solar-like stars \citep{leanetal1995}. It was concluded that TSI during the Maunder minimum was about 3--4 W/m$^2$ less than at present  which translates into a solar radiative forcing\footnote{Solar radiative forcing  is a direct energy source to the Earth and is related to the change in TSI  by $\Delta F=\Delta{\rm TSI} \cdot (1-A)/4$, where A is the Earth's albedo.}   $\Delta F_{\rm P-M}$$\sim\,$0.5--0.7 W/m$^2$.  However,  these results were not  confirmed by large surveys of solar-like stars and are no longer considered to be correct \cite{halllockwood2004}. Recent reconstructions based on the magnetic field surface distribution \citep{wangetal2005,krivovaetal2007} and on extrapolation of the  assumed correlation between TSI and the open magnetic flux \citep{lockwoodetal1999} during the last three minima \citep{steinhilberetal2009,frohlich2009}  resulted in a low solar forcing value within the range  $\Delta F_{\rm P-M}\approx$ 0.1--0.2 W/m$^2$.

\section{Effects of  solar radiative  forcing on climate}
Variations on time-scales up to the 27-day rotational period have an important influence on space weather but not on terrestrial climate. The effects of the 11-year solar cycle are clearly detected in the atmosphere and are widely discussed in the literature \citep[e.g.,][]{egorovaetal2004, haigh2007}, while the imprints from variations on longer time-scales are more subtle 
{\citep{grayetal2010}}. Analysis of  historical data suggests a strong correlation between solar activity and natural climate variations on centennial time-scales, such as the colder climate during the Maunder (about 1650--1700 AD) and Dalton (about 1800--1820 AD) minima as well as climate warming during the steady increase in solar activity in the first half of the twentieth-century \citep{siscoe1978,hoytandschatten1997, solomonetal2007, grayetal2010}. Numerous attempts to confirm these correlations based on different climate models have shown that it is only possible if either the applied perturbations of direct solar radiative forcing are large (consistent with a direct solar radiative forcing from the present to Maunder minimum  $\Delta F_{\rm P-M}$$\sim\,$0.6--0.8 W/m$^2$) or the amplification of a weak direct solar forcing is substantial. Because the majority of recent $\Delta F_{\rm P-M}$ estimates (see Sect. 1) are only in the range 0.1--0.2 W/m$^2$, and amplification processes have not been identified, the role of  solar forcing in natural climate change  remains  highly uncertain \citep{solomonetal2007}. {  In this paper we show that the solar forcing may be significantly larger than reported in the recent publications. }

\section{Methods}\label{sec:methods}
%\subsection{General concept}\label{sec:GC}
In this paper we present a new alternative technique, avoiding calibration of our model with presently observed TSI variations and extrapolation to the past. We assert that the amount of magnetic energy that  remains present \citep{deWijnetal2009} at the surface of a spotless (i.e. quiet) Sun is the main driver of solar irradiance variability on centennial time scales. The main concept of our technique is to determine the level of the magnetically enhanced contribution to the irradiance of 
the present quiet Sun.
{  Then, for the reconstruction to the past, this magnetically 
enhanced component has to be scaled with the  proxies for the quiet Sun activity.  However a proxy for the long-term activity of the quiet Sun does not yet exist. A good candidate for such a proxy is the small-scale turbulent magnetic fields accessed with the Hanle effect \citep{stenflo1982}. However the consecutive measurements of these fields are limited to the last few years \citep{kleint2010}. Therefore, we assume that the existing proxies of the solar activity averaged over the two solar cycles period can also describe the activity of the quiet Sun.  Averaging of proxies allows sufficient time for the magnetic components to decay to the quiet network, and then to even smaller magnetic features as the decay process can take up to several years \citep{solankietal2000}.  In other words we set the small-scale activity  (which defines the fractional contributions of quiet Sun components) to be proportional to the large-scale activity. 
We want to state clearly  that this proposition  is an assumption, which is however  in  line with a high-resolution observations of the present Sun.
The large-scale structure due to strong magnetically active features is repeated on a smaller scale in less active regions, and even in the apparently quietest areas there is still a mosaic of regions of different magnetic field strengths, reminiscent of fractal structure \citep{deWijnetal2009}.  }

Hence, the time-dependent irradiance $ I_{\rm quiet}(\lambda,t)$ of the quiet Sun in our reconstruction can be calculated as
\begin{equation}
\frac{I_{\rm quiet}(\lambda,t) - I_{\rm min.state}(\lambda)}{\mathcal <Proxy>_{22} \left ( t  \right )}  =  \frac{I_{\rm quiet} \left (\lambda, {t_0} \right) - I_{\rm min.state}(\lambda)} { {{\mathcal <Proxy>_{22}} \left (t_0 \right)}  } ,  \nonumber 
\end{equation}
where  $ I_{\rm quiet}(\lambda,t)$ is the time dependent  irradiance of the quiet Sun and $t_0$ denotes a reference time. {  ${\mathcal Proxy}_{22}$ is the averaged over a 22-year period value  of the proxy for the solar activity.  The 22-year period was chosen because  the cosmogenic isotope data used for the reconstruction (see below) are available as 22-year averaged data \citep{steinhilberetal2008}. } We have set 1996 as the reference year, i.e. $I_{\rm quiet}(\lambda,t_0)$ is the irradiance of the quiet Sun as  observed during the 1996  minimum. {  Let us notice that  the quiet Sun irradiance was roughly constant for the last 3 cycles (see below). Therefore the solar spectrum during the 1996 minimum is a good representation of the present quiet Sun  spectrum.} $ I_{\rm min.state}(\lambda)$ is the irradiance of an absolute minimum state of the Sun, 
with a minimum of remaining magnetic flux emerging on the solar surface. Thus, the term in brackets can be considered as the enhancement level of the present quiet Sun with respect to the Sun in its most inactive state.

The prominent, readily observable active regions on the Sun also 
contribute to the  variability in irradiance. Thus, the full 
solar variability is described by:
\begin{equation}
I(\lambda,t) \equiv  I_{\rm quiet}(\lambda,t) + I_{\rm active}(\lambda,t),
\end{equation}
where $ I_{\rm active}(\lambda,t)$ is the contribution to solar irradiance from active regions e.g. sunspots, plages, and network.
 $I_{\rm active}(\lambda,t)$ is calculated following the approach by \citet{krivovaetal2003} (see also the online Sect.~\ref{sec:SS}) {  and, as calculations are done with annular resolution, is proportional to the sunspot number. The group sunspot number used in our reconstruction is taken from NOAA data center and described in \citet{hoytschatten1998}}.

{  The term  $ I_{\rm active}(\lambda,t)$ in the Eq. (2) describes the cyclic component of solar variability due to the 11-year activity cycle, while the slower  long-term changes due to the evolution of the small-scale magnetic flux are given by the Eq. (1). $ I_{\rm active}(\lambda,t)$ term can only be calculated for the time when sunspot number is available and therefore our reconstruction has 22-year resolution for the Holocene and annual resolution from 1610 to present.} 

The choice of the model for the minimum state of the Sun is a crucial point in our technique as it  defines the amplitude of the reconstructed solar irradiance variability. Observations of the Sun with relatively high spatial resolution show that the present quiet Sun  is still highly inhomogeneous even though it appears spotless. Measurements obtained with the Harvard spectroheliometer aboard Skylab were used  to derive brightness components of the quiet Sun and to construct corresponding semi-empirical solar atmosphere structures \citep{vernazzaetal1981, fontenlaetal1999}. The darkest regions with  the least amount of magnetic flux corresponds to the faint supergranule cell interior (component A). This component comes closest to describing the most inactive state of the Sun.
Therefore, in our approach  we set
$ I_{\rm min.state}(\lambda) \equiv  I_A(\lambda) $. Thus,   the basic magnitude of solar irradiance variations is given by
the difference between the irradiance of the present quiet Sun (composed from a distribution of brightness components defined in supporting online material) and the 
irradiance from component A (see Eq. (1)).

For the reconstruction to the past this amplitude is scaled with proxies for solar activity. Two proxies are available for the reconstruction: Group sunspot number, which is available from the present to 1610 AD, and the solar modulation potential extending back to circa 7300 BC. The latter is a measure of the heliospheric shielding  from cosmic rays  derived  from the analysis of cosmogenic isotope abundances in tree rings or ice cores,
and is available with a time resolution of 2-3 solar cycles \citep{steinhilberetal2008}.  Although sunspot number  dropped to zero  for a long time during the Maunder minimum, the solar cycle was uninterrupted \citep{beeretal1998, usoskinetal2001}  and the modulation potential did not  fall to zero.  Hence, a reconstruction based solely  on  sunspot number  may underestimate the solar activity during the Maunder minimum. {  Therefore in our reconstruction we used the solar modulation potential to  calculate the long-term variations  and sunspot number to  superpose them  with  the 11-year cycle variations  (see  the Online Section 6.2). }

\begin{figure}
\resizebox{70mm}{!}{\includegraphics[]{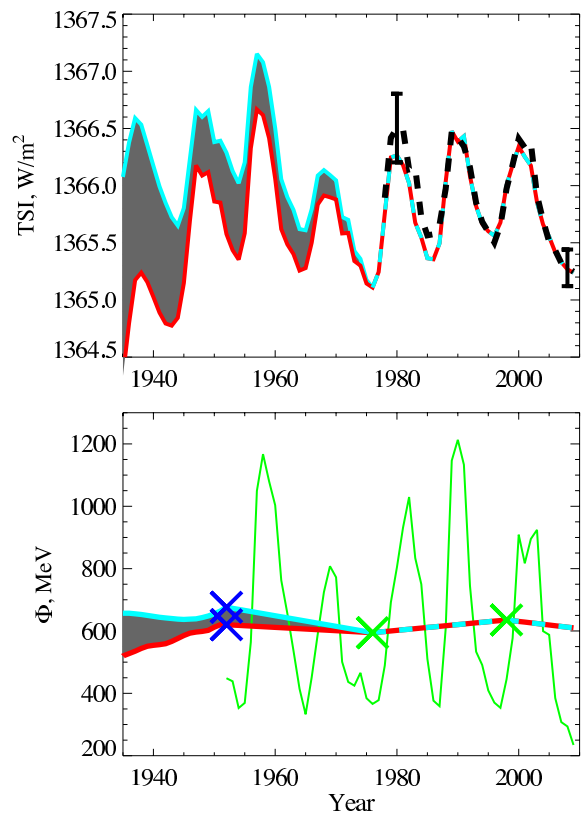}}
\caption{Modulation potential and TSI reconstruction for the last 70 years. Lower panel: Yearly averaged neutron monitor data (green) and two modulation potential composites (red and cyan curves, see the discussion in the text).  
Upper panel: TSI reconstructions based on the two modulation potential composites (red and cyan curves). The black dashed line is the observed TSI from the PMOD composite. Error bar for 1980 corresponds to 1/4  the difference between two published TSI composites, and the error for 2008 is taken from \citet{frohlich2009}. The reconstructed TSI curves are normalized to the 1996 minimum and the grey-shaded region indicates the intrinsic uncertainty due to differences in the modulation potential data.}
\label{fig:LTEmol}
\end{figure}

\begin{figure*}
\centering
\includegraphics{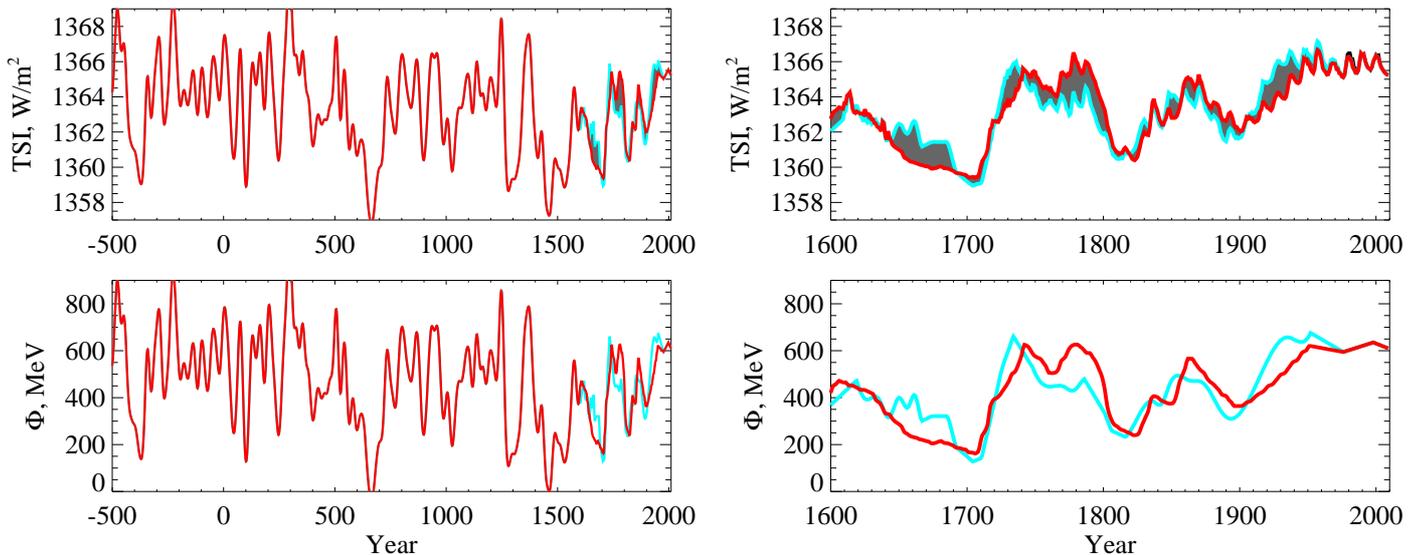}
\caption{Modulation potential (lower panel) and TSI reconstructions (upper panel) for the last 2500 years. Data prior to 1600 AD are based on the modulation potential derived from $^{10}$Be records from the Greenland Ice core Project (red curves). Data since 1600 AD are based on the two composites shown in Fig. 1 (red and cyan curves). The grey-shaded area indicates the intrinsic uncertainty.}
\end{figure*}

\begin{figure*}
\centering
\includegraphics{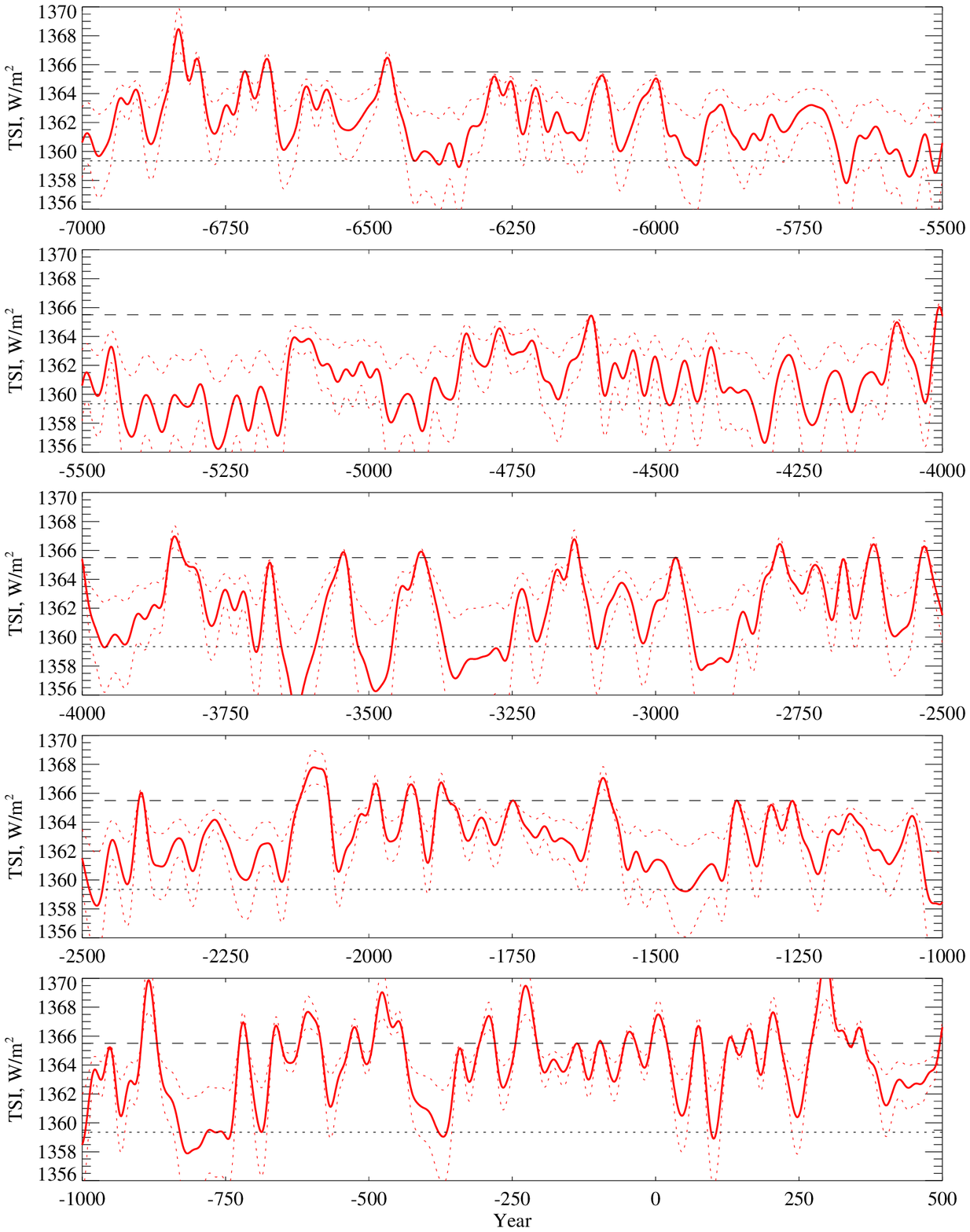}
\caption{TSI reconstruction from 7000 BC to 500 AD. The reconstruction is based on the modulation potential derived from $^{10}$Be records from the Greenland Ice core Project. {  Red dotted lines indicate the estimated error bars of the reconstruction.} Black dashed and dotted lines indicate  TSI  for the 1996 solar minimum and the lowest TSI   during the Maunder minimum, respectively. }
\end{figure*}

{  The modulation potential used in the calculations is based on the composite of  data determined from the cosmogenic isotope records of $^{10}$Be and neutron monitor.  $^{10}$Be data are available up to about 1970 \citep{McCrackenetal2004} and neutron monitor data, which are used to calculate the current solar modulation potential, are available since the 1950s. Three different datasets of the $^{10}$Be data were used: measurements at DYE 3, Greenland, and the South Pole provided by \citet{McCrackenetal2004}  for the reconstruction back to the Maunder minimum, and measurements from the Greenland Ice core Project provided by \citet{vonmoosetal2006} for the reconstruction back to 7300 BC. The neutron monitor data were provided by   \citet{usoskinetal2005a}. Although they  are available with a monthly resolution,  we calculated the 22-year mean of the neutron monitor in order to homogenize the data sets and because only averaged modulation potential can be used as a proxy for the quiet Sun activity. The transition from $^{10}$Be to neutron monitor data  is shown in the lower panel of Fig.~1.   Prior to 1952 the  composites are based on $^{10}$Be data (cyan based on  DYE 3, and red on  South Pole records, both with 22-year resolution). Values between 1952 and 1998 are obtained by linearly interpolating the following consecutive data points: two blue crosses at 1952 (the last data points used from  $^{10}$Be records),  green crosses at 1976 and 1998 (22-year averages of the neutron monitor data).
Values after 1998 require  knowledge of the next 22-year average (between 2010 and 2031) of the modulation potential. 
We assume this average to be 92\% of the previous average (between 1988 and 2009), which allows us to reproduce the observed TSI  minimum in 2008.The datasets mentioned above were derived using different assumptions of the Local Interstellar Spectra (LIS). To homogenize the data we converted the modulation potential from \citet{McCrackenetal2004}  and \citet{usoskinetal2005a}  to LIS by \citet{castagnoliandLal1980} which is used in \citet{vonmoosetal2006}. For conversion we  applied  a  method suggested by \citet{steinhilberetal2008}.  \citet{herbstetal2010} analyzed the dependency of the modulation potential on the applied LIS models and show that negative values of the modulation potential in  \citet{vonmoosetal2006} data can be corrected with another LIS model.  Let us notice that the solar forcing in our reconstruction is determined by the relative values of the modulation potential so change to the different LIS model will introduce only several percents correction, which is much less than estimated accuracy of the reconstruction.  {In our dataset the value of the reference modulation potential $ {\mathcal <Proxy>_{22}} \left (t_0 \right)$ (see Eq. (1)) is equal to 631 MeV. The relative error is estimated to be less than 10\% \citep{usoskinetal2005a}. } 
}

Synthetic solar spectra are calculated with a state-of-the-art radiative transfer code \citep{haberreiter2008, shapiroetal2010}. We discuss the calculations in more detail in online Sect.~\ref{sec:SS}, where in addition we show that the  concept of quiet Sun activity scaling can also be expressed in terms of varying fractional contributions from different components of the quiet Sun.

\section{Results and Discussion}
In the upper panels of Fig.~1 and  Fig.~2  we present the TSI reconstructions, {  which are obtained after the integration of the Eqs. (1) and (2) over the wavelengths and normalization of the quiet Sun value for the reference year 1996 to 1365.5 W/m$^2$} .  As the sunspot number is only available since 1610 AD the reconstruction of the full solar cycle variability with an annual resolution extends back only 400 years. Both reconstructions in the right-hand panels of Fig.~2 are based on the $^{10}$Be data sets mentioned above. {The difference in the reconstructions   allows   the error originating from the uncertainties in the proxy data to be estimated (20-50 \% {  in the solar forcing value}, depending on the year).  This is large but still significantly less than the change in irradiance between the present and the Maunder minimum. }  {Both reconstructions suggest a significant increase in TSI during the  first half of the twentieth-century as well as low solar irradiance during the Maunder and Dalton minima.} The difference between the current and reconstructed TSI during the Maunder minimum is about $6\pm 3$ W/m$^2$ (equivalent to a solar forcing of $\Delta F_{\rm P-M}$$\sim\,$1.0$\pm0.5$ W/m$^2$) which is substantially larger than recent estimates (see Sect.~\ref{sec:intro}).  Note that as our technique uses 22-year means of the solar modulation potential our approach cannot be tested with the last, unusual solar minimum in 2008.  In order to reproduce the current minimum as shown in Fig.~1 we have adopted a value of  584 MeV for the future 22-year average in 2020 (which is 92\% of the 22-year average for 1988--2009).   

The reconstruction before 1500 AD in the left-hand panel of Fig.~2   (which stops for clarity at 500 BC)  is based on  $^{10}$Be records from the Greenland Ice core Project 
\citep{vonmoosetal2006}.  The modulation potential during the Maunder minimum is about 3--4 times less than at present but not zero. However, it has decreased to zero several times in the past and the corresponding TSI was even smaller than during the Maunder minimum. There were also several periods when the modulation potential, and hence TSI, were higher than the present value. {   The reconstruction back to 7000 BC is presented in  Fig.~3. {The choice of model A introduces an uncertainty of the order of 30\%,  which is estimated by comparing model A to  other possible candidates for the minimum state of the quiet Sun, e.g., model B from \citet{vernazzaetal1981}.  Combining this with the uncertainties of the proxy data outlined in Fig.~2 we can roughly estimate the uncertainty  of our solar forcing value to be 50\%.   In additional to the $^{10}$Be-based data of solar activity there are several $^{14}$C-based datasets \citep[e.g.,][]{solanki2004N, vonmoosetal2006, muscheler2007, usoskinLR}.  Employment of these datasets will lead to a somewhat different values of the solar variability, which is, however, covered by our rough order of magnitude estimate of the overall uncertainty of the reconstruction.}
  }

\begin{figure}
\resizebox{84mm}{!}{\includegraphics{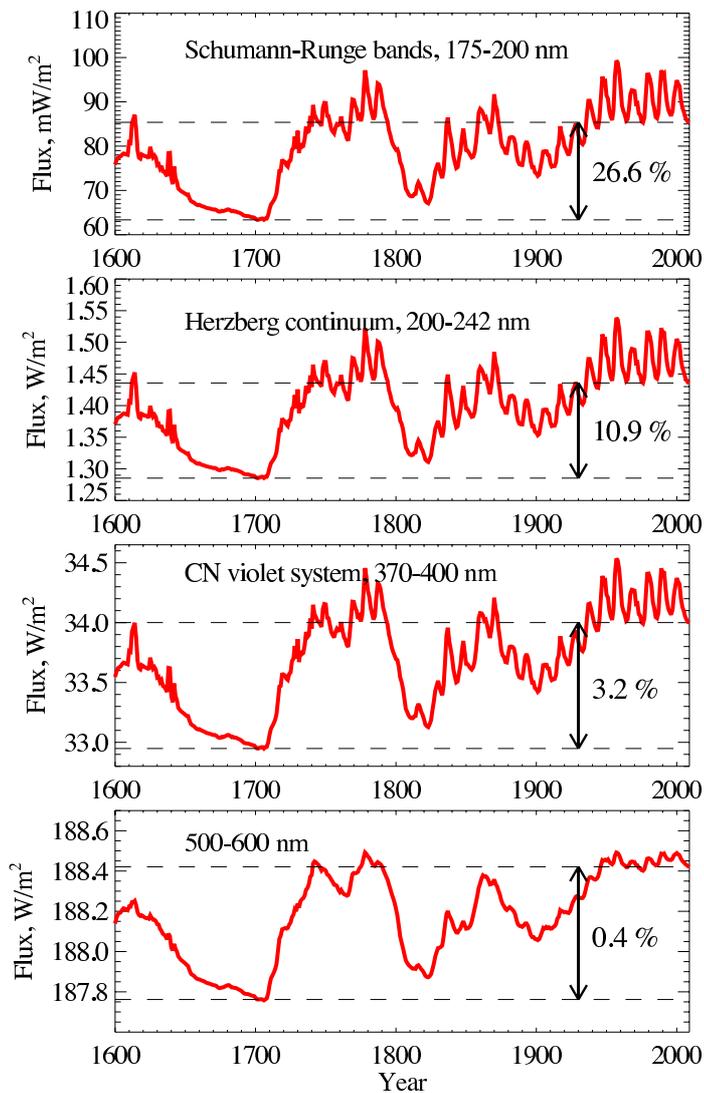}}
\caption{Reconstruction of the integrated spectral irradiance in selected wavelength bands.  This reconstruction is based on the composite of the $^{10}$Be  South Pole record and neutron monitor data (red curve in Fig. 2).  Panels from bottom to top: The 500--600 nm band is representative of the time evolution of the visible irradiance. The CN violet system is one of the most variable bands accessible from the ground. The Herzberg continuum  is crucial for ozone production in the stratosphere. The Schumann-Runge band is important for  heating processes in the middle atmosphere.}
\end{figure}

Our TSI reconstructions give  a value of  $\sim$1 W/m$^2$ per decade for   the period 1900--1950.
The Smithsonian Astrophysical Observatory (SAO) has a  32-year record of ground-based observations  for 1920--1952. 
Although the SAO data are disputed in reliability \citep{SAO} and clearly contain a non-solar signal they are the only available long-term measurements of TSI  in the first half of the twentieth-century.  The data show an increase of 1$\pm 0.5$  (1.5$\pm 0.5$) W/m$^2$  per decade for the period 1928--1947 (1920--1952).
 As we are aware that maintaining a stable calibration to better than 0.1 \% over 30 years  is very demanding we cannot
claim that the historical data confirm our reconstruction. Nevertheless, it is intriguing to note how well  the SAO trend agrees
with our TSI reconstruction.

Our reconstructed solar spectral irradiance comprises spectra from 130 {nm} to \linebreak 10  $\mathrm{\mu}$m.  Fig.~4 presents a reconstruction of the integrated flux for several, selected spectral regions. The contrast between different brightness components of the quiet Sun is especially high in the UV, which results in a large historical variability of the UV spectral irradiance.   The irradiance in the Schumann-Runge bands and Herzberg continuum increases from the Maunder minimum to the present  by about 26.6\% and 10.9\% respectively, which is much larger than  0.4\% for TSI and the visible region.
The variability is also relatively high around the CN violet system whose strength is very sensitive to even small temperature differences due to the high value of the dissociation potential. 
The large UV variability reported here is especially of importance to the climate community because it influences
 climate via an indirect, non-linearly amplified forcing \citep{haigh1994,egorovaetal2004}.

We are aware that the choice of model A is responsible for a relatively large fraction of the uncertainty in our results.
Higher resolution observations have recently become available, and model A could possibly be improved in future studies.  We emphasize that model A is not the coldest possible quiet Sun model and therefore our estimate is not a lower limit of  the Sun's energy output. The coldest model would be a non-magnetic atmospheric structure without a chromosphere and corona. 
As model A contains some remaining magnetic activity our approach does not imply that the solar dynamo stops during the periods when the modulation potential is equal to zero. Let us also notice that the modulation potential never reached zero for the last 400 years (see Fig.~2).

\section{Conclusions}
{  

We present a new technique to reconstruct total and spectral solar irradiance over the Holocene.
We obtained a large historical solar forcing between the Maunder minimum and the present, as well as a significant increase in solar irradiance in the first half of the twentieth-century.
Our value of the historical solar forcing is remarkably larger than other estimations published in the recent literature.

We note that our conclusions can not be tested on the basis of the last 30 years of solar observations because, according to the proxy data, the Sun was in a maximum plato state in its {\it  {long-term}} evolution. 
All recently published reconstructions agree  well during the satellite observational period and diverge only in the past.
This implies that  observational data do not allow to select and favor one of the proposed reconstructions. Therefore, until new evidence become available we are in a situation that different approaches and hypothesis yield different solar forcing values. {Our result allows the climate community  to evaluate the full range of present uncertainty in solar forcing.}

 The full dataset of the solar spectral irradiance back to 7000 BC is available upon request.

 }

\begin{acknowledgements}
We are grateful  to Friedhelm Steinhilber and J\"urg Beer  for useful discussions and help with data.
The research leading to this paper was supported by the Swiss National Science Foundation under grant CRSI122-130642(FUPSOL) and also  received funding from European Community's Seventh Framework Programme (FP7/2007-2013) under
grant agreement N 218816 (SOTERIA). M.H. appreciates funding from the Swiss Holcim Foundation.
\end{acknowledgements}

%\bibliographystyle{aa}
%\bibliography{shapiro}

\Online
\section{Spectral synthesis}{\label{sec:SS}}
\subsection{Quiet Sun}
The quiet Sun is a combination of different brightness components and the evolution of their fractional contributions drives its activity and long-term irradiance variability. 
The four main components of the quiet Sun \citep{vernazzaetal1981,fontenlaetal1999} are: component A (faint supergranule cell interior), component  C (average supergranule cell interior), component  E (average network or quiet network), and component F (bright network).

We calculated the synthetic spectra $I_A$, $I_C$, $I_E$, $I_P$  of all these components employing the NLTE (non-local thermodynamic equilibrium)  COde for Solar Irradiance \citep{haberreiter2008, shapiroetal2010} (COSI). Recently,  \citet{shapiroetal2010}  showed that  COSI calculations with the atmosphere model for component C reproduces spectral irradiance measurements from the last two solar minima with good accuracy.  This is used in Eq. (1) from the main text, where  $ I_{\rm present}(\lambda)$ is substituted by the  irradiance  $I_C(\lambda)$ for component C.

The evolution of the magnetic activity of the quiet Sun can be represented by the time-dependent fractional contributions (i.e. filling factors) of different components of the quiet Sun. The quiet Sun can be described by model A with varying contributions from the brighter components. 
We  compose the quiet Sun using model A, and model E which is an adequate representation of the brighter contributions.    

Using COSI we demonstrate (see Fig.~5) that the solar irradiance for model C and, therefore, measured solar irradiance for the last two minima,  can be successfully reproduced with a combination of $\alpha_{\rm present}^A=43\%$  model A and  $\alpha_{\rm present}^E=57\%$ model E ($\alpha_{\rm present}^A+\alpha_{\rm present}^E=1$):
\begin{equation}
I_{\rm quiet}^{\rm present}(\lambda) = I_C (\lambda)=\alpha_{\rm present}^A I_A (\lambda) + \alpha_{\rm present}^E I_E(\lambda). 
\end{equation}

The linear scaling of the magnetic activity  of the quiet Sun with proxies, as described in the main text, is equivalent to setting the model E filling factor to be proportional to a chosen proxy. So that the time-dependent irradiance of the quiet Sun can be calculated:
\begin{equation}
I_{\rm quiet}(\lambda, t)= \left (1-\alpha^E \left (  t  \right )  \right ) I_A(\lambda)+\alpha^E(t) I_E(\lambda),
\end{equation}
where 
\begin{equation}
\alpha^E(t)=\frac{\mathcal Proxy(t)}{\mathcal Proxy_{\rm present}}  \alpha_{\rm present}^E.
\end{equation}
The set of Eqs. (2-4) is equivalent to Eq. (1).

There is an ongoing discussion of whether the trend in the filling factor of the quiet network  can be detected. It was suggested \citep{foukalmilano2001} that the analysis of  the historical Mt. Wilson observations exclude the existence of the trend between 1914 and 1996. However, this analysis was criticized for using uncalibrated data \citep{solankikrivova2004} and also contradicts  other studies \citep{lockwood2010}.
Thus only the future long-term monitoring of the quiet Sun with high resolution or recently proposed \citep{kleint2010} monitoring of weak turbulent magnetic fields can help to clarify this question.

\subsection{Active Sun}
From 1610 onward we have additional information from sunspot number, which allows the calculation of the active regions contribution to the solar irradiance ($I_{\rm active} (\lambda,t)$ in Eq. (2)). For this we follow the approach by \citet{krivovaetal2003}. Because our main goal is to  reproduce centennial solar variability and because magnetograms are unavailable for historical time periods, we scale the faculae and the active network filling factors with the sunspot number  instead of using filling factors derived from available magnetogram data. The synthetic spectra are then added according to their filling factors, and TSI is then determined by integrating over all wavelengths.

\begin{figure}
\resizebox{84mm}{!}{\includegraphics{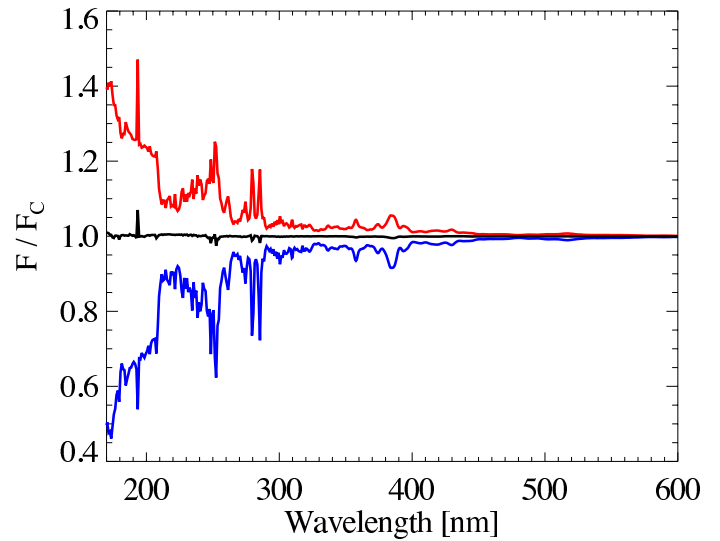}}
\caption{Representation of the quiet Sun by a combination of models. Ratios of synthetic solar spectra of  models A (blue) and E (red) to  model C. The black line results by combining  57\%  of the model E spectrum with 43\% of the model A spectrum. The spectral signatures are almost perfectly cancelled, implying  that the combination of both spectra is equal to the model C spectrum. The latter  has been demonstrated to accurately reproduce the observed solar spectra during the last two minima\citep{shapiroetal2010}. Hence,  a combination of spectra for models A and E reproduces the observed quiet Sun spectrum. }
\label{fig:conc}
\end{figure}

\end{document}